\documentclass{article}

\usepackage{graphics,latexsym}
\usepackage{graphicx}
\usepackage{amsmath, amssymb,amsthm}

\begin{document}

\title{High Frequency Electrical Oscillations in Cavities}
\author{Daniele Funaro}

\date{}
\maketitle

\centerline{Dipartimento di Fisica, Informatica e Matematica} 
\centerline{Universit\`a di Modena e Reggio Emilia}
\centerline{Via Campi 213/B, 41125 Modena (Italy)} 
\centerline{daniele.funaro@unimore.it}

 \begin{abstract}
If the interior of a conducting cavity (such as a capacitor or a coaxial cable) is supplied with a very 
high-frequency electric signal, the information between the walls propagates with an appreciable delay,
due to the finiteness of the speed of light. The configuration is typical of cavities having size larger
than the wavelength of the injected signal.
Such a non rare situation, in practice,  may cause a break down of the performances of the device.
We show that the classical Coulomb's law and Maxwell's equations do not correctly
predict this behavior. Therefore,  we provide
an extension of the modeling equations that allows for a more reliable determination
of the electromagnetic field during the evolution process.
The main issue is that, even in vacuum (no dielectric inside the device), the fast variation
of the signal produces sinks and sources in the electric field, giving rise to zones
where the divergence is not zero. These regions are well balanced, so that their average
in the domain is zero. However, this behavior escapes the usual treatment with 
classical electromagnetism. 
\end{abstract}

\section{Introduction}

This paper concerns with the modeling of the behavior of capacitors (or other analogous devices) under
high-frequency stimulation. By this we mean that the distance of the armatures
is of the order of the wave-length of the injected signal. At these regimes, due to the finiteness of the
velocity of propagation of the electromagnetic signals, one cannot account on
Coulomb's law, since this requires an infinite speed of propagation. Nevertheless, serious 
contraindications also arise in the framework of
classical Maxwell's equations. The theoretical analysis conducted in Ref.~\cite{debr} shows
that, even in very elementary cases, the full set of Maxwell's equations may bring to
incompatibility problems. Indeed, the enforcement of the wave equation for the electric field 
in conjunction with the divergence-free condition, may enter in conflict with boundary conditions
in a very wide range of circumstances. 
\par\smallskip

The example discussed in Ref.~\cite{debr}, though elementary, is representative of other
more serious situations that will be discussed later on. We briefly recall the basic setting.
We assume that our capacitor is made of
two equal plates of shape $\Omega$. These are parallel and placed at a
distance $d$. The vertical direction coincides with the $z$-axis and the
plates are situated at the positions $z=0$ and $z=d$. We assume to
be in presence of an ideal capacitor, i.e., the electric
field stays perpendicular to each plate surface. Moreover,  the
charge can be uniformly modified on the plates. A more realistic
case is the one treated in  Ref.~\cite{feyn}, v.2, chapter 23,
where the signal is injected at points centered on the plates.
Scrupulous examination of this setting may turn out to be rather involved,
however. Other extensions may be taken into consideration;
they just make the study far more complicated, while we would like
here to catch the spirit of the underlying problem.
\par\smallskip

The two armatures are subjected to a time-varying forcing source $\alpha (t)$,
with $\alpha (0)=0$.
As far as the  boundary conditions are concerned, we
assume that, on both plates, one has for any $t>0$:
\begin{equation}\label{conv}
{\bf E }~=~(0,0,\alpha (t)/d) 
\end{equation}
\par\smallskip
In addition, we require that laterally the capacitor is totally insulated.
This amounts to say that the electric field ${\bf E}=(E_x,E_y,E_z)$ is orthogonal to the
normal ${\bf n} = (n_x,n_y,n_z)$ $= (n_x,n_y, 0)$ to the boundary $\partial\Omega\times [0,d]$.
Note that $n_x$ and $n_y$ do not depend on $z$.
It is usual to couple this boundary relation with the following one: ${\bf
B}\times {\bf n}=0$, $\forall~t$. For an overview on boundary conditions one may check
for instance  Ref.~\cite{cai}; Ref.~\cite{born}, p. 4; Ref.~\cite{monk}, p. 8; Ref.~\cite{ryl}, 
p. 4; as well as many other texts.
The last condition can be differentiated in
time, obtaining $(\partial {\bf B}/\partial t)\times {\bf n}=0$,
$\forall~t$, which is translated in terms of the curl of ${\bf E}$ by the Faraday's law. 
Therefore, the full set of boundary constraints for the lateral surface of the
capacitor may be written in function of the electric field as:
\begin{equation}\label{rote}
{\bf E}\cdot{\bf n}~=~0 ~~~~~~~~~~{\rm curl}{\bf E}\times{\bf n}~=~0
~~~~~{\rm on}~\partial\Omega\times [0,d]~~~~~\forall ~t
\end{equation}
This means that, for the three components of ${\bf E}$, one has the three boundary equations:
\begin{equation}\label{rotec}
E_xn_x+E_yn_y=0~~~~~~~~~\frac{\partial E_x}{\partial y}=
\frac{\partial E_y}{\partial x}~~~~~~~~~\frac{\partial E_z}{\partial
x}n_x+\frac{\partial E_z}{\partial  y}n_y=0
\end{equation}

The discussion is at the moment academic, so we will not question ourselves
on the mechanism that allows for the creation of a uniform difference of potential between
the surfaces of the capacitor. 
The reader that does not feel comfortable with these conditions of total insulation may
always argue with a device having very large plates, a small gap, and subject to a fast variation of charge. 
The setting is mainly mathematical and concerns with the well-posedness 
of a boundary-value problem for a well-known set of PDEs. 
The question is how to distinguish between correct or incorrect boundary conditions, when
such a distinction is not a priori coded in the model, but only relies on experience.
In this introduction, the aim is to instill a mathematical suspect and leave for later
the examination of more realistic devices.
\par\smallskip

For the classical Maxwell's system, we have to solve
the wave equation for the electric field, i.e.:
\begin{equation}\label{onde}
\frac{\partial^2 {\bf E}}{\partial t^2}~=
~c^2 \Delta {\bf E}
\end{equation} 
Here $c$ is the speed of light in vacuum.
Equation (\ref{onde}) is well-posed under the boundary constraints
suggested in (\ref{rotec}). One easily discovers that $E_x=E_y=0$ inside the capacitor.
In fact, by setting $E=E_z$, where $E$ does not depend on $x$ and $y$, 
the following one-dimensional wave-equation admits unique solution (under suitable initial guesses):
\begin{equation}\label{wavez}
\frac{\partial^2 E}{\partial t^2}~=~c^2
\frac{\partial^2 E}{\partial z^2}~=~c^2 \Delta E
\end{equation} 
and it is fully compatible with  (\ref{rotec}). Again, due to the uniqueness property,
this is enough to conclude that the vector ${\bf E}$ must be of the form $(E_x, E_y, E_z)=(0,0,E)$ 
and solve the vector wave-equation in (\ref{onde}) together with (\ref{rote}).
It is not difficult to verify that the so obtained ${\bf E}$ cannot
have zero divergence (i.e.: ${\rm div}{\bf E} =\partial E
/\partial z\not =0$), since by (\ref{conv}) we should have $E(z)=\alpha(t)/d$ for any $0\leq z\leq d$ and
this is not compatible with the wave equation. 
Hence, we find ourselves in a very general situation ($\Omega$, $d$ and $\alpha$ are arbitrary),
in which the set of equations turns out to be overdetermined. This is true independently of
the displacement of the magnetic field (actually, not considered here). 
This says that, whatever
is the applied  $\alpha \not =0$, one always finds discrepancies by putting together
the following three ingredients: wave equation, divergence-free condition, boundary conditions.
As one introduces asymmetries in the capacitor, we cannot avoid creation of magnetic field, so
that an analysis similar to the one given above becomes extremely difficult from a mathematical
viewpoint. By the way, we can expect similar incongruences. As a matter of fact, it would be rather strange (and disappointing) 
to discover that the only bad situations occurs under the hypothesis of symmetry, although such a
perfection is not actually encountered in real life. This exception would imply that some passages to the limit
are somehow forbidden. The study given in  Ref.~\cite{debr} provides more insight on this controversial issue.
Further results pointing out inconsistencies in the Maxwell's model, at certain regimes, can be found in 
Ref.~\cite{eng2},  \cite{fun2}.
\par\smallskip

We conclude this introduction by saying that the field ${\bf E}$ obtained by requiring 
${\rm div}{\bf E}=0$ is acceptable for slow-varying $\alpha $, but becomes unrealistic
when the velocity of variation of $\alpha$ is comparable with $c$. If $\alpha (t)=\sin c\omega t$ is 
a sinusoidal signal of wave-length $\lambda = 2\pi /\omega$, significant discrepancies
are expected when $\lambda\approx d$. The reason is that the change of information
at one boundary is not communicated in time to the other boundary, and vice versa. 
The only chance to get a better description of what
actually happens is to get rid of equation ${\rm div}{\bf E}=0$. We see how to do this in the coming section. 
 
\par\medskip
\section{Extension of the Maxwell's model}

We briefly describe an extension of the equations modeling electromagnetism, as it has been introduced in
Ref.~\cite{fun2}. Such a generalization overcomes the troubles that may emerge in the 
application of the classical Maxwell's model, as for example in the case pointed out in the introduction.
\par\smallskip

As usual ${\bf E}$ and ${\bf B}$ denote the electric and magnetic field, respectively.
We set by definition: $\rho ={\rm div}{\bf E}$. 
In addition, we have a new velocity field ${\bf V}$, which has the role
of indicating the direction and speed of development of the energy
flow (a sort of generalization of the Poynting vector).
The new set of equations reads as follow:
\begin{equation}\label{sfem2}
\frac{\partial {\bf E}}{\partial t}~=~ c^2{\rm curl} {\bf B}~
-~\rho {\bf V}
\end{equation}
\begin{equation}\label{sfbm2}
\frac{\partial {\bf B}}{\partial t}~=~ -{\rm curl} {\bf E}
\end{equation}

\begin{equation}\label{sfdb2}
{\rm div}{\bf B} ~=~0
\end{equation}

\begin{equation}\label{slor2}
\rho\left( \frac{D{\bf V}}{Dt}~+~\mu ({\bf E}+{\bf V}\times {\bf
B})\right) ~=~-\nabla p
\end{equation}

\begin{equation}\label{presse}
 \frac{\partial p}{\partial t} ~=~\mu\rho ({\bf E}\cdot {\bf V})
\end{equation}
with $D{\bf V}/Dt=\partial {\bf V}/\partial t+({\bf
V}\cdot \nabla ){\bf V}$.
Equation (\ref{sfem2}) turns out to be the Amp\`ere
law for a flowing immaterial current with density $\rho$,
naturally associated with the movement of the electromagnetic wave.
Note that condition $\rho\not =0$ may happen to be true also in vacuum. 
Relation (\ref{slor2}) is the \index{Euler's equation}
Euler's equation for inviscid fluids with a forcing term of
electromagnetic type given by the vector ${\bf E}+{\bf V}\times
{\bf B}$, which recalls the Lorentz's law. 
Up to dimensional scaling, the scalar $p$ plays the role of pressure
(explanations about the meaning of $p$ will be provided in section 3). 
The constant $\mu$ has the dimension of charge divided by mass. In
Ref.~\cite{fun2}, $\mu$ has been estimated to be approximately equal to
$2.85\times 10^{11}~{\rm Coulomb / Kg}$. Relation (\ref{presse}) says that 
pressure may raise as a consequence of a lack of
orthogonality between ${\bf E}$ and ${\bf V}$. Let us finally observe that it is not possible 
to recover the wave equations for  ${\bf E}$ and  ${\bf B}$ in the new context
(unless $\rho =0$). 
\par\smallskip

The revised set of equations is similar to that ruling plasma physics (see, e.g., Ref.~\cite{jackson},
chapter 10), with the difference that no matter is present in our case.
We trivially return to the  Maxwell's setting in vacuum by imposing $\rho =0$
and $p=0$. The important fact here is that the vector field  ${\bf V}$ in
(\ref{sfem2}), (\ref{slor2}), (\ref{presse}) is unknown and it is not directly
related to the product  ${\bf E}\times {\bf B}$, as suggested by the classical theory.
\par\smallskip

Since a sort of current term is present in (\ref{sfem2}), due to conservation arguments, a
continuity equation must hold.
This does not need to be imposed independently, since it is easily
obtained by taking the divergence of (\ref{sfem2}):
\begin{equation}\label{cont}
\frac{\partial \rho}{\partial t}=\frac{\partial}{\partial t}{\rm div} {\bf E}=
{\rm div}\left( c^2{\rm curl} {\bf B}-\rho {\bf V}  \right)=
-{\rm div}(\rho {\bf V})
\end{equation}

By denoting with ${\bf E}_s$ the solenoidal part of ${\bf E}$, the three equations
(\ref{sfem2}), (\ref{sfbm2}) and (\ref{cont}), describe in unique way the evolution
of the triplet ${\bf E}_s$, ${\bf B}$, $\rho$. This is coherent with the standard Maxwell's model
in vacuum where ${\bf E}_s={\bf E}$ and $\rho =0$ (so equation (\ref{cont}) disappears).

\par\smallskip

The above modeling equations hold in vacuum, but they can easily
take into account the presence of external sources, such as charges and currents due to
electrified bodies. It is enough to add the proper source terms, as for instance
the usual current field ${\bf j}$ on the right-hand side of (\ref{sfem2}).
The corresponding new versions 
of the  Amp\`ere law and continuity equations are:
\begin{equation}\label{sfem2cu}
\frac{\partial {\bf E}}{\partial t}= c^2{\rm curl} {\bf B}-\rho {\bf V}-{\bf j}
\qquad\qquad
\frac{\partial \rho}{\partial t}
+{\rm div}(\rho {\bf V})+{\rm div}{\bf j}=0
\end{equation}
The last is the standard continuity equation if one eliminates the middle term.
The above relations control at the same
time both the contribution given by the forcing term and the presence
of a spontaneous non-vanishing divergence of ${\bf E}$ inside the pure electromagnetic phenomenon.
Note that $\rho$ does not distinguish among these two contributions. The quantity $\rho$ is always the
divergence of the electric field by definition, thus, as before, it does not constitute an equation by itself.
If ${\bf j}\not =0$, then $\rho$ is
not zero, nevertheless $\rho$ can be different from zero even if
${\bf j} =0$, and this is the real novelty of the approach.
\par\smallskip

For instance, in the framework of plasma physics, one could in principle describe,
using a unique set of equations, the evolution of electrons and ions (contained in the term
${\bf j}$ through the coupling with some transport equation, such as Vlasov's), together with the exchange
of electromagnetic radiation between these charges. It is like to work with the movement of three
species (negative charges, positive charges, pure electromagnetism).
Considering that particles in a plasma usually develop at very high speed, this
generalization may turn out to be extremely useful.

\par\smallskip

A particular case of primary interest is when $D{\bf V}/Dt =0$ and
$p=0$, where, according to Ref.~\cite{fun2}, the solutions have been named {\sl free-waves}.
In this situation one has (\ref{sfem2}), (\ref{sfbm2}), (\ref{sfdb2}),
plus the (geometric) relation:
\begin{equation}\label{lore}
{\bf E}+{\bf V}\times {\bf B}=0
\end{equation}
that indicates that no ``forces'' are acting on the electromagnetic wave. Examples
of free solutions are the following ones.
\par\smallskip

In Cartesian coordinates, by orienting the electric field along the $z$-axis, a
full solution of the set of equations (\ref{sfem2}), (\ref{sfbm2}), (\ref{sfdb2}),
(\ref{lore}) is:
\begin{equation}\label{sol1}
{\bf E}=\Big(0,~0,~cf(z)g(ct-x)\Big), \ 
{\bf B}=\Big(0,~-f(z)g(ct-x), ~0\Big), \ {\bf V}=(c,~0,~0)
\end{equation}
where $f$ and $g$ can be arbitrary. This wave shifts at the
speed of light along the $x$-axis. Note that it is possible to
enforce the condition $\rho =0$ (thus, returning to the set of
Maxwell's equations) only if $f$ is a constant function (i.e., an entire plane wave, displaying
infinite energy). This first example shows that the new solution space turns out to be
rather large. Waves like those given in (\ref{sol1}) are also available in cylindrical
coordinates $(r,\phi ,z )$. For instance, we may consider:
\begin{equation}\label{sol3}
{\bf E}=\Big( cf(r)g(ct-z),~ 0,~0\Big),\
{\bf B}=\Big(0,~f(r)g(ct-z), ~0\Big),\ {\bf V}=(0,~0,~c)
\end{equation}
where now the shifting is in the direction of the $z$-axis. The function $f$ must tend to zero for
$r\rightarrow 0$, in order to be able to define the fields
on the $z$-axis.  Starting from this
setting there are no solutions (except zero) in the Maxwellian
case. If $f$ and $g$ have compact
support, one obtains a kind of bullet traveling undisturbed at
the speed of light. This opens the way to a classical description
of photons. The reader can imagine the impact that this discovery has on the
discussions on {\sl wave-particle duality} (comments in this direction 
are given in Ref.~\cite{fun4}). 
\par\smallskip

Finally, going into spherical coordinates $(r, \theta, \phi )$, one can
analyze waves associated to perfect
spherical fronts. We have:
\begin{equation}\label{sol4}
{\bf E}=\Big(0,~\frac{c f(\theta )}{r}g(ct-r),~0\Big),\
{\bf B}=\Big(0,~0,~
\frac{f(\theta )}{r}g(ct-r)\Big), \ {\bf V}=(c,~0,~0)
\end{equation}
The electric and magnetic fields are tangent to spherical surfaces
and shift in the radial direction. Note that $f$ must tend to zero at the poles, i.e., when 
$\theta \rightarrow 0$ and $\theta \rightarrow \pi$. Let us observe that there are no waves of this type in the
Maxwellian case (i.e., with ${\rm div}{\bf E}=0$). Note in fact that, in the famous
Hertzian solution (see, e.g., Ref.~\cite{jackson}, p. 394), the Poynting vectors are not of radial type.
Other general sets of solutions are provided in the appendix of Ref.~\cite{fun4}.
\par\medskip

\section{The charge of a capacitor}

We now discuss the model problem regarding the capacitor suggested in the introduction.
We recall that this is just a first step to validate the model on a
very simple situation. Later we will face upgraded examples.
\par\smallskip

We rewrite all the equations (\ref{sfem2}), (\ref{sfbm2}), (\ref{sfdb2}),
(\ref{slor2}), (\ref{presse}) by assuming that ${\bf B}=0$,
${\bf E}=(0,0,E)$, ${\bf V}=(0,0,V)$, where $E$ and $V$ do
not depend on $x$ and $y$. We get:
\begin{equation}\label{pri}
\frac{\partial E}{\partial t}~=~-V\frac{\partial E}{\partial z} 
\end{equation}
\begin{equation}\label{sec}
\frac{\partial E}{\partial z}\left( \frac{\partial V}{\partial t}
~+~ V\frac{\partial V}{\partial z}~+~\mu E\right)
~=~-\frac{\partial p}{\partial z}
\end{equation}
\begin{equation}\label{ter}
\frac{\partial p}{\partial t}~=~\mu E V\frac{\partial E}{\partial z} 
\end{equation}
From (\ref{pri}) and (\ref{ter}), one recovers the pressure (up to additive constants):
\begin{equation}\label{ter2}
\frac{\partial p}{\partial t}~=~-\mu E \frac{\partial E}{\partial t} 
~=~-\frac{\mu}{2} \frac{\partial E^2}{\partial t} ~~~~~~\Rightarrow~~~~~~
p~=~-\frac{\mu}{2}E^2
\end{equation}

Afterwards, substituting $p=-(\mu /2)E^2$ in (\ref{sec}), we arrive at the nonlinear
inviscid transport equation:
\begin{equation}\label{bur}
\frac{\partial V}{\partial t}~+~ V\frac{\partial V}{\partial z}~=~0
\end{equation}
By symmetry arguments, it is natural to assume that both the lower plate ($z=0$)
and the upper one ($z=d$) are inflow boundaries, so that $V=c$ at $z=0$ and
$V=-c$ at $z=d$. When the capacitor is discharged we have $V=0$ for $0<z<d$.
Thus, the solution of (\ref{bur}) is a shock-wave:
\begin{equation}\label{burso}
V=
\begin{cases}
c & {\rm if} ~ 0\leq z < ct \cr 0 & {\rm if} ~ ct\leq z\leq d-ct
\cr -c & {\rm if} ~d-ct < z\leq d 
\end{cases}
\end{equation}
while $t<d/2c$. As $t$ reaches the value $d/2c$, we have a stationary discontinuous
step centered at $z=d/2$. Equation (\ref{pri}) has to 
be intended {\sl almost everywhere}. It does not hold for
instance at the point $d/2$ where the shock develops. 
\par\smallskip

In order to discuss a specific case, we propose an experiment similar to that presented in Ref.~\cite{debr}. 
The difference of potential
between the plates is increased quadratically in a given interval
of time, after which is kept constant. In this way, the information propagates from
the boundaries to the interior. The initial datum is $E=0$ for $t=0$.
\par\smallskip

To obtain an approximate solution, we
apply a classical explicit first-order upwind scheme. 
The interval $[0,d]$ is divided in $n$ equal parts ($n$ even) of size
$\Delta z$. The time-step $\Delta t$ is less than $\Delta z/c$ in order
to satisfy the CFL condition. We denote by $\eta^k_i$ the approximation
of $E$ at the node $z_i$ at the instant $t_k$. The initial datum (capacitor
discharged) implies $\eta^0_i=0$, for all $i$. Successively, we proceed
as follows:
\begin{equation}\label{schem}
\eta_i^{k+1}=
\begin{cases}
\eta^k_i-\frac{\Delta t}{\Delta z} v^k_i (\eta^k_i -\eta^k_{i-1}) & 
{\rm for} ~i=1,...,n/2-1 
\cr\cr \frac{1}{2}(\eta^k_{i+1}+\eta^k_{i-1}) & {\rm for} ~ i=n/2
\cr\cr \eta^k_i-\frac{\Delta t}{\Delta z} v^k_i (\eta^k_{i+1} -\eta^k_i) & 
{\rm for} ~i=n/2+1,...,n-1 
\end{cases}
\end{equation}
Note that the coupling of (\ref{pri}) and (\ref{bur}) is nonlinear,
so that the two equations must be advanced together in time.
At the boundaries we have: $\eta_0^{k+1}=\eta_n^{k+1}=\alpha (t_{k+1})/d$.
According to (\ref{burso}), in (\ref{schem}) we defined
for $ ck\Delta t <d/2$:
\begin{equation}\label{vdis}
v^k_i=
\begin{cases}
c & {\rm if} ~~ i\Delta z < ck\Delta t 
\cr 0 & {\rm if} ~~ ck\Delta t \leq i\Delta z \leq d-ck\Delta t 
\cr -c & {\rm if} ~~  d-ck\Delta t < i\Delta z
\end{cases}
\end{equation}
For $i=n/2$, in (\ref{schem}) we imposed that $\eta_i^{k+1}$ is
an average of the neighboring values at the previous step. This condition
has no direct physical meaning, by the way it becomes irrelevant as $\Delta z \rightarrow 0$.
\par\smallskip

In Fig. 1 we give the results of an experiment where $d=1$, $c=1$, $\alpha (t)=t^2$
if $t<1$ and $\alpha (t)=1$ if $t\geq 1$.
We recall that $E=E_z$ denotes the vertical component of the electric
field; the figure shows its behavior at various time steps.
The information propagates from the boundaries (the two plates) 
and reaches the interior with some delay. At time $t=1$ the increase
of $\alpha$ is stopped. After a while, the shape of $E$
becomes perfectly constant. This means that the capacitor is fully
charged and that the distribution of $E$ between the plates is uniform, as predicted
by the Coulomb's law (${\rm curl}{\bf E}=0$ and ${\rm div}{\bf E}=0$). 
The duration of the process if finite and not exponentially asymptotic, 
as roughly obtained from the study of LRC circuits. Note that, 
due to the nonlinearity of the model equations,
the superposition principle does not hold during the time evolution.
The reader can see schematically the phases of the charging process in Fig. 2. 
\par\smallskip

\begin{center}
\vspace{-.1cm}
\begin{figure}[h]
\centerline{\includegraphics[width=6.5cm,height=6.cm]{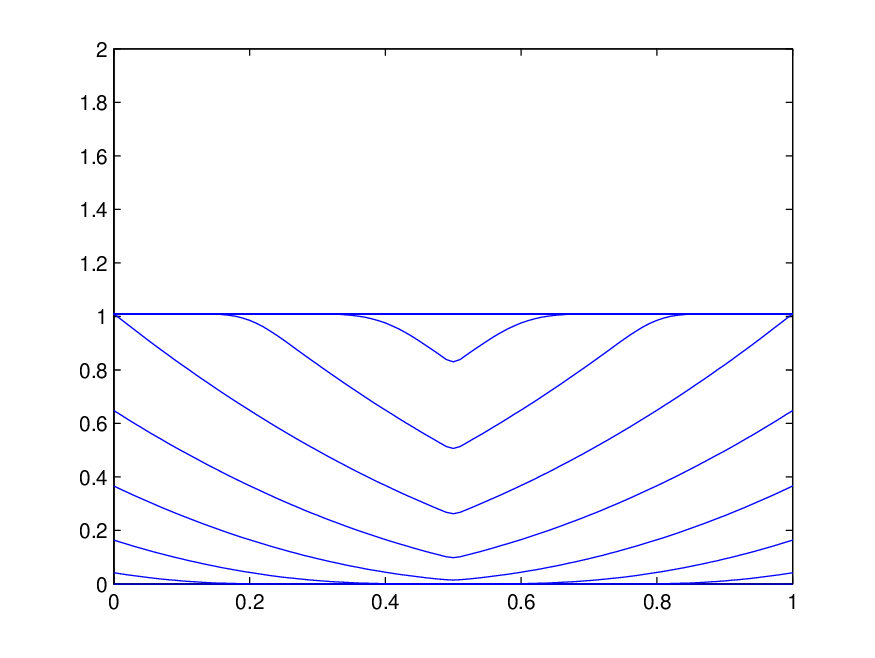}
\hspace{-.2cm}\includegraphics[width=6.5cm,height=6.cm]{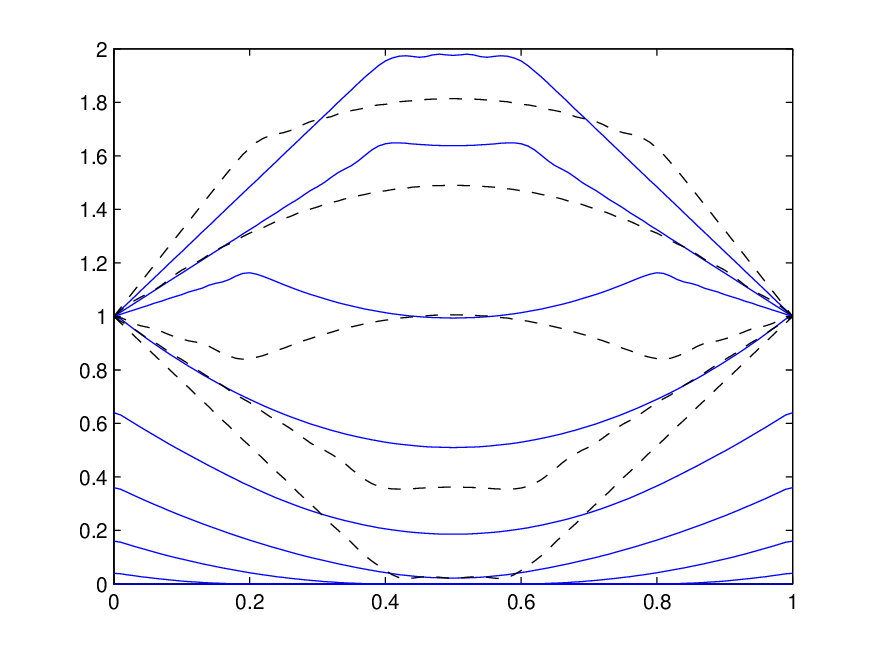}}
\vspace{-.1cm}
\begin{caption}{\small Behavior with respect to time (left) of the function
$E=E_z$ in (\ref{pri}) with $V$ obtained from (\ref{bur}). This is compared
(right) with the solution of the wave equation (\ref{wavez}) corresponding to the
same boundary data. Unreliable oscillations are produced in the second case.
The initial transient, where the distribution monotonically grows (solid lines), is followed 
by a periodic regime (dashed lines). The principal modes ($\sin \pi z$ and $\sin 3\pi z$) 
are clearly visible.}
\end{caption}
\end{figure}
\end{center}\vspace{-.1cm}

Going back to Fig. 1, we have compared the above results with those obtained by solving
numerically the wave equation (\ref{wavez}) with the same boundary data (see, e.g., Ref.~\cite{inan} or Ref.~\cite{ryl}). 
Here, being the equation second-order in time, together with the initial condition $E=0$, we added 
an initial guess on the time derivative, i.e.: $\partial E /\partial t =0$ at $t=0$.
The results obtained are unacceptable from the physical viewpoint. As we stop to increase
the boundary data, the inner electric field continues to develop and assumes an
oscillating behavior that never damps. The intensity 
reaches values that are larger than those attained at the boundaries.
In addition,  almost all the displacements have divergence different from zero, in
contrast with the classical requirements. Note instead that the condition
$\rho ={\rm div}{\bf E}=0$ has been removed from the modified model.
\par\smallskip

We provide a quick explanation for the different performances of the two models.
The usual Amp\`ere's law in vacuum ($\rho =0$) requires that the generic vector
$\partial {\bf E}/\partial t$ is the {\sl curl} of another vector, but this is not always 
true, especially when certain boundary constraints are involved. 
In fact, the new version of the equations includes on the right-hand side of the Amp\`ere's law the
term $\rho {\bf V}$ which is not solenoidal. Despite the presence of $\rho {\bf V}$,
the equation of continuity (\ref{cont}) still holds and ensures
that, apart from the boundaries, there are no forcing current terms 
inside the capacitor. Note that the integral
of $\rho$ on the whole volume of the capacitor is zero, in accordance with Gauss theorem. The model allows for the
momentary creation of internal electric sinks and sources, that are zero in average at any time $t$ and disappear
when the capacitor is fully charged.
\par\smallskip

\begin{figure}[h]
\begin{center}
\centerline{\includegraphics[width=12.cm,height=7.5cm]{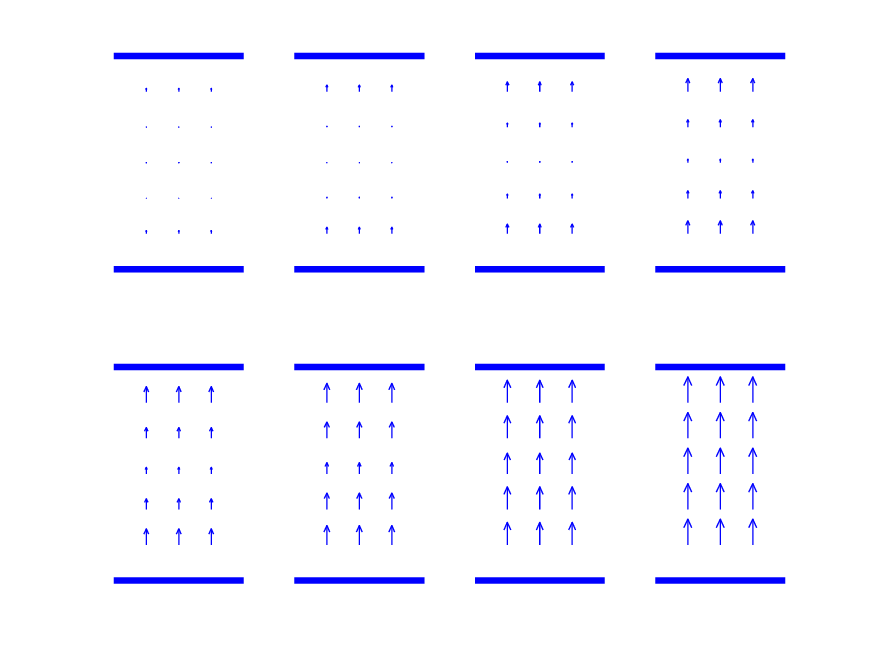}}
\vspace{-.3cm}
\begin{caption}{\small Displacement of the electric field ${\bf E}$ during the charge
of a capacitor. The divergence of ${\bf E}$ remains different from zero, until 
the process is terminated.}
\end{caption}
\end{center}
\end{figure}

For the solution of Fig. 1 (left), we can evaluate the scalar $p$ according to (\ref{ter}),
assuming that the distribution at time  
$t=0$ is $p=0$. Although the explicit expression of $p$ is available (see (\ref{ter2})),
the integration of (\ref{ter}) allows for the recovering of the unknown additive constant.
As the charging process is completed there is a uniform negative distribution of $p$ inside the capacitor.
If we attribute to $p$ the meaning of a fluid pressure (up to multiplicative
dimensional constant), this result is interesting since it may explain
the Coulomb's attraction of the two charged plates in terms of mechanical forces.
The merging of electrodynamical and mechanical effects is one of the major 
features of the modified model, allowing for instance the study of the acceleration 
imparted to very small material bodies as a consequence of pressure exerted
by photons. This is for instance what happens in {\sl optical tweezers}. Some computations in this direction are
provided in Ref.~\cite{funka}. 
\par\smallskip

It is wise to add some comments about the role of potentials in our analysis. 
Condition  (\ref{conv}) should be interpreted as a consequence of the fact that the parallel plates
of the capacitor are equal and uniformly charged. The concept of ``difference of potential'' 
descends from Coulomb's law and it is not appropriate in a time-dependent situation.
In this case, one needs to introduce the scalar potential $\Phi$ and the vector potential ${\bf A}$
as usual: ${\bf B}={\rm curl}{\bf A}/c$, ${\bf E}=-\nabla\Phi -(\partial {\bf A}/\partial t)/c$
(see, e.g., Ref.~\cite{jackson}, p. 219).
Moreover, we suppose to be in the Lorenz gauge: $(\partial\Phi /\partial t)+c\hspace{.1cm}{\rm div}{\bf A}=0$.
In the specific case we are dealing with, one can set  ${\bf A}=(0,0,A)$ and require
that both $\Phi$ and $A$ only depend on $z$. In this fashion, one soon obtains ${\bf B}=0$.
Passing through (\ref{pri}), with little manipulation, one arrives at the equality (see
also Ref.~\cite{fun2}, section 2.5):
\begin{equation}\label{dwa}
c\left(\frac{\partial^2 A}{\partial t^2}-c^2\frac{\partial^2 A}{\partial z^2}\right)
=V\left(\frac{\partial^2 \Phi}{\partial t^2}-c^2\frac{\partial^2 \Phi}{\partial z^2}\right)
\end{equation}
The above relation is milder than that imposed by Maxwell's equations, where both of the
terms of the equation must vanish. This explains why the new model is able
to describe a wider range of problems.
\par\smallskip

All the examples analyzed so far are based on boundary conditions that force the
magnetic field ${\bf B}$ to stay zero during the evolution. It is evident that
this is just a ``toy problem" on which one can validate the set of equations
from the mathematical viewpoint.
In real phenomena, a magnetic field actually develops (see, e.g., Ref.~\cite{bart}), so that
the model must be used in full without simplification (an elementary example
with  ${\bf B}\not =0$ will be given in section 4).
The theoretical analysis of what happens during the charge of a capacitor
becomes prohibitive, since simple solutions, as the ones examined
here, are no longer available. Therefore, a further study may require serious
numerical computations. Nevertheless, the aim here is to illustrate with elementary facts
that the revision of the model is justified by some basic applications.

\par\smallskip

A spontaneous question is how a commercial software for electromagnetics 
(such as Ref.~\cite{cst}) behaves when dealing with the charge of
a capacitor. The answer is that numerical codes have their foundation on
a strong theoretical background, which is available for a restricted range
of problems, that do not include, of course, the option we are examining here.
Therefore, there is no way to input the necessary data, because these do not belong to the
configuration supported by the code. The most straightforward approach is to work
in the so called {\sl time-frequency domain}. This consists in replacing
time-derivatives through multiplication by the term $i\omega$
(see, e.g., Ref.~\cite{monk}, section 1.2), and solve the wave equation by looking 
for eigenvalues of the Laplace operator. Boundary and initial conditions ensure existence
and uniqueness, but, as we checked here, the additional constraints on divergence may turn out to be
incompatible. To bypass this difficulty, the software improperly restricts the 
number of boundary conditions to be imposed (we recall that the theory establishes that at each point of the
boundary six conditions are necessary, three for the electric field and three for the
magnetic one). This is done surreptitiously by forcing the so called ``natural conditions",
that in general correspond to Neumann type constraints and that are more easily adaptable to the
divergence-free relation. These hidden conditions do not usually appear in the {\sl menu} bar.
\par\smallskip

\begin{center}
\begin{figure}[h]
\vspace{-.8cm}
\centerline{\includegraphics[width=12.cm,height=11.3cm]{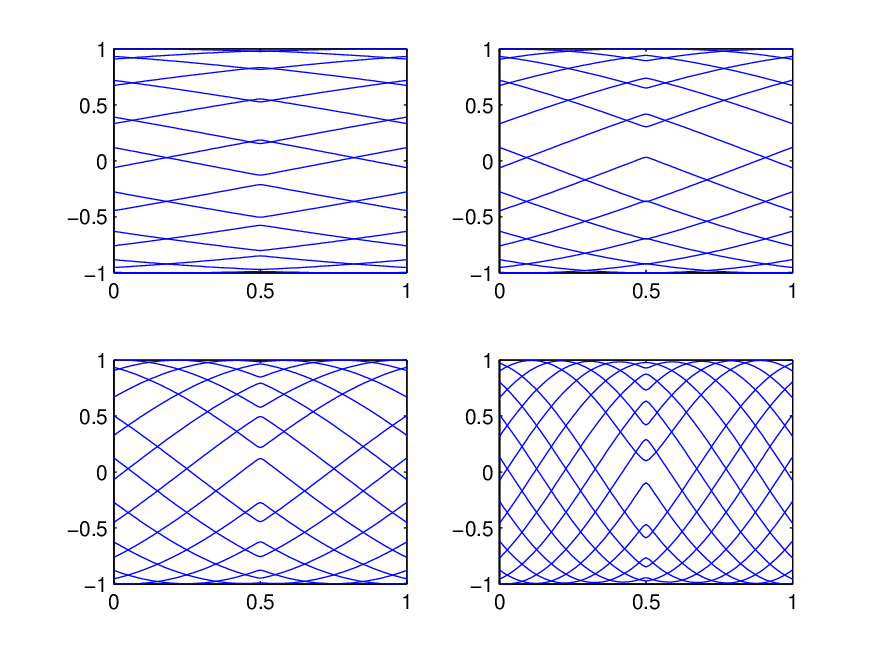}}
\vspace{-.5cm}
\begin{caption}{\small Behavior with respect to time of the function
$E$ in (\ref{prire}) at different values of $t$ and four choices
of the parameter $\omega$, namely: .5, 1, 2 ,4.}
\end{caption}
\end{figure}
\end{center}

As a further test, we propose to compute the distribution of the electric field
${\bf E}=(0,0,E)$ inside the capacitor, when the difference of potential is a sinusoid 
function $\alpha (t)=\sin c\omega t$. We neglected the initial transient and plot 
the solutions after a certain given time, in order to stabilize the internal oscillations (see Fig. 3,
corresponding to $d=1$, $c=1$ and different values of the parameter $\omega$). 
This means that $V=c$ in the interval $[0,d/2[$ and $V=-c$ in the interval $]d/2, d]$.
Thus,  (\ref{pri}) reduces to the transport equations:
\begin{equation}\label{prir}
\frac{\partial E}{\partial t}+c\frac{\partial E}{\partial z}=0 ~~~~{\rm in}~[0,d/2[,
~~~~~~~~
\frac{\partial E}{\partial t}-c\frac{\partial E}{\partial z}=0 ~~~~{\rm in}~]d/2, d]
\end{equation}
The two branches of solutions (one departing from the lower plate and the other from the upper one) 
meet at the central point $z=d/2$.
The global distribution is continuous but not smooth. 
In this special circumstance, based on the boundary conditions, it is easy to provide an explicit 
expression for $E$:
\begin{equation}\label{prire}
E=\sin \omega (ct -z) ~{\rm if}~ z\in [0,d/2[, ~~~
E=\sin \omega (ct +z -d) ~{\rm if}~ z\in ]d/2, d]
\end{equation}
\par\smallskip

As expected, the signal departing from the armatures of the capacitor arrives with some delay
at the center. The asynchrony depends on the speed $c$ of propagation of the signal
(to be adjusted if one is not in vacuum), the distance $d$, and the frequency $c\omega /2\pi$
of the applied source. At any prescribed time $t$, for $\omega$ very high, the electric field 
may even change sign several times within the gap $]0,d[$. As a consequence, when the above values are 
critical, the capacitor
may not work according to the usual rules. This property is well-known and, if a dielectric
is present, it is attributed to the inability of the microscopic internal dipoles
to change polarization in the requested time. Here, we are noting that this can be also true in 
pure void. The most basic law says that {\sl reactance} is
inversely proportional to frequency. Therefore, at high frequencies, the capacitor is like
short-circuited, though this crude explanation is not sufficient to understand what is  effectively  
going on at its interior.
The modified model predicts what can actually happen at these extreme regimes.
On the contrary, the classical approach, not only is unable to model a reasonable development of the
fields, but leads to inconsistent conclusions (see again Fig. 1, right).

\par\medskip
\section{High-frequency fields in other devices}

In order to show that we are able to handle more realistic
situations, we present here the results of further developments regarding
the behavior of an electromagnetic field between two conductors working
at high-frequency.
\par\smallskip

We begin with observing that, for $0\leq \gamma \leq 1$, the following
electric and magnetic fields (expressed in Cartesian coordinates),
solve the entire set of equations  (\ref{sfem2}), (\ref{sfbm2}), (\ref{sfdb2}),
(\ref{slor2}), (\ref{presse}):
\begin{equation}\label{prireg}
{\bf E}=\Big(0, ~0, ~ c\sin \omega (ct -\gamma x-\beta z)\Big) \qquad
{\bf B}=\Big(0,  ~ -\gamma \sin \omega (ct -\gamma x-\beta z), ~0\Big)
\end{equation}
with $\beta =\sqrt{1-\gamma^2}$ and ${\bf V}=c(\gamma , 0, \beta)$, so that
the velocity of propagation is that of light: $\Vert {\bf V}\Vert =c$. 
Therefore, this time ${\bf B}$ has the chance
to be different from zero. 
For $\gamma =1$ we have a classical plane wave of frequency $c\omega /2\pi$.
In this last case, the intensity of ${\bf E}$ is $c$ times that of ${\bf B}$
and the fields are fully transverse (the Poynting vector is lined up with ${\bf V}$).
For $\gamma =0$ we are in a case similar to
(\ref{prire}), where ${\bf B}=0$ and ${\bf E}$ is parallel to ${\bf V}$. 
For $0<\gamma <1$, we find intermediate situations. The divergence $\rho$ takes
the values $-c\beta\omega \cos \omega (ct -\gamma x-\beta z)\not =0$, so that we are not in
the usual Maxwell's farmework in vacuum.
\par\smallskip

As in the previous section, one can study what happens inside a cavity with conductive walls.
This time the armatures are subject to a dynamical constraint proportional to $\sin \omega (ct -\gamma x)$.
The same happens for instance on the conducting surfaces of a coaxial cable
(see Ref.~\cite{cheng}, chapter 8; Ref.~\cite{jackson}, chapter 8; Ref.~\cite{grif}, section 9.5.3).
Actually, the solutions in (\ref{prireg}) can be easily translated in cylindrical
coordinates (see also (\ref{sol3})). In this way, for $\gamma =1$, we rediscover the typical 
{\sl TE mode} configuration (see the last picture of Fig. 4). 
\par\smallskip

\vspace{.3cm}
\begin{center}
\begin{figure}[h]
\vspace{-.1cm}
\centerline{\includegraphics[width=19.cm,height=4.cm]{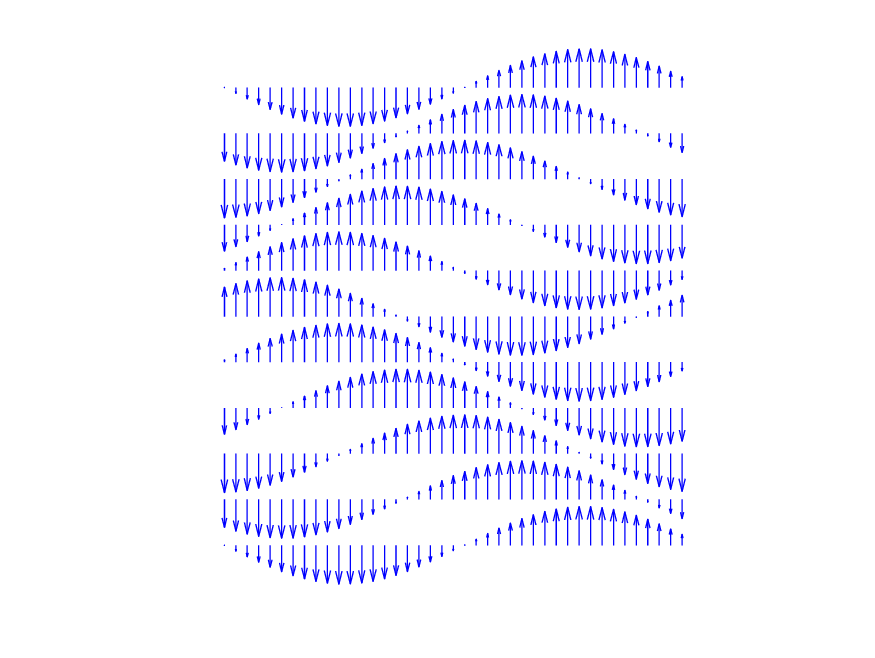}}
\vspace{-.4cm}
\centerline{\includegraphics[width=19.cm,height=4.cm]{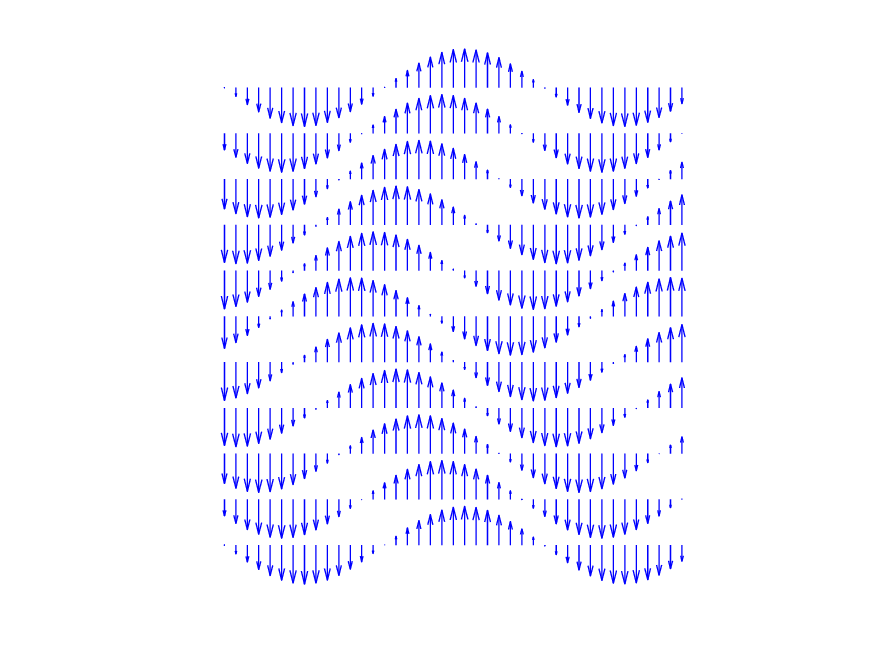}}
\vspace{-.4cm}
\centerline{\includegraphics[width=19.cm,height=4.cm]{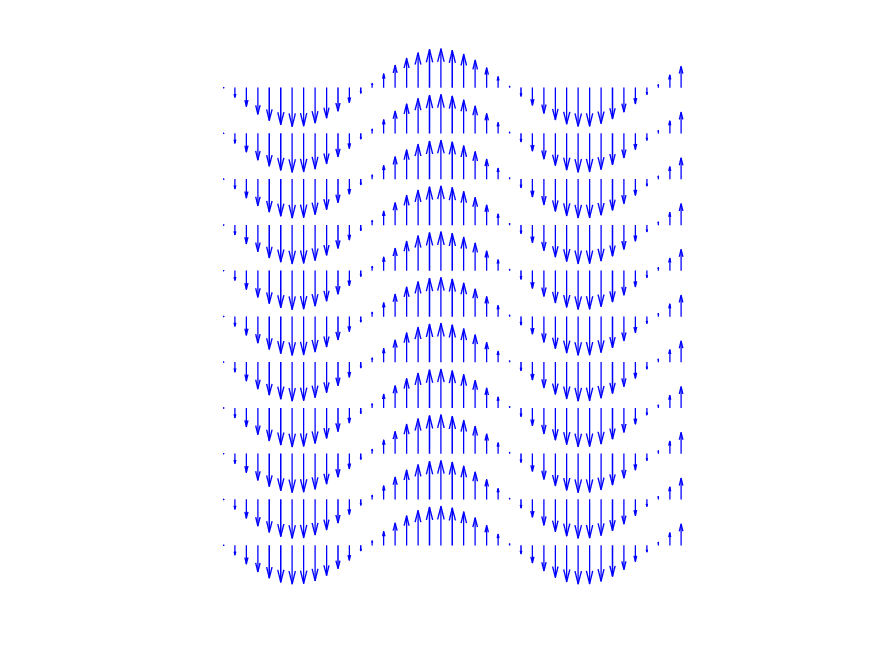}}
\vspace{-.1cm}
\begin{caption}{\small Displacement of the electric field according to
(\ref{prireg}) for a given $\omega $, when $\gamma =0.6$, $\gamma =0.9$ and $\gamma =1$,
respectively. The non-stationary boundary conditions are the 
same on the lower and upper plates, but the information reaches with a delay
(depending on $\gamma$) the center of the capacitor.}
\end{caption}
\end{figure}
\end{center}

With the help of our extension we can study the case
when the change occurring at the boundaries is extremely fast. In this fashion,
together with the usual drift in the direction of ${\bf E}\times {\bf B}$, we also
have a transverse drift due to the time needed to carry the information from
the boundaries to the center of the device. Some examples relative to a parallel
plate capacitor (of large extension and small gap $d$) are documented in Fig. 4.
Of course, here the frequency must be very high (small wave-length in comparison to the vertical dimensions).
It is known that, at these conditions, a coaxial cable displays heavy power losses.
{\sl Attenuation } in waveguides is studied for example in Ref.~\cite{cheng}, p. 409. 
Transmission in rectangular pipes is examined in Ref.~\cite{grif}, p. 410, through a repetitive
bouncing of the traveling wave against the walls. This is different from what we developed here,
but the two things do not exclude each other, so that the effects may be combined.
\par\smallskip

The final example we would like to mention is quite crucial and concerns with the study
of the near-field in antenna emissions. Whatever is its shape, an  
antenna is commonly subject to frequencies whose wave-length is of the same order of magnitude
of the device. Thus, a complete analysis should take care of the delay necessary
to communicate the electromagnetic information from the conductors towards the gap
between them. As we saw, this study requires an extension of the Maxwell's
model in vacuum, since it seems natural to ask the divergence $\rho$ to be different
from zero (see also Ref.~\cite{fun3}). The spherical wave solution provided in (\ref{sol4})
is already a step ahead in this investigation, because, for any given function $f\not=0$,
the divergence of the electric field does not vanish (with the exception of the singular
case: $f(\theta )=1/\sin (\theta )$). However, this displacement still does not 
incorporate the possibility of simulating the requested delay, while it can be
certainly ascribed to the far-field for suitable choices of the function $f$.
\par\smallskip

Significant experiments on near-field propagation have been reported in Ref.~\cite{khol1}
and Ref.~\cite{khol2}. There, a registered anomalous behavior of the signal suggests 
a local superluminar regime. Although the (partial) conclusions of the authors are not in line with  
the work presented here, those results are interesting for a couple of reasons.
First of all, in Ref.~\cite{khol3}, a theoretical explanation is developed, where a
suitable mechanical tensor is added to the electromagnetic stress tensor.
Such an approach is very similar to the one we are proposing here (see also Ref.~\cite{fun2}, chapter 4).
The mass tensor contains, under the form of momentum, the velocity of the
 system ${\bf V}$, which is not necessarily equal to $c$.
 Secondly, the model here discussed is able to include explicit cases when $\vert {\bf V}\vert \not =c$
 (Ref.~\cite{fun2}, chapter 5). 
Thus, the above mentioned experiments may be a good battleground to further justify the new model.
Prototypes of new antennas, inspired from similar principles, have been documented in Ref.~\cite{hart}.
 The implementation of numerical
techniques, in conjunction with the model here discussed, may certainly be a valid aid for
the analysis of the functioning mechanism of antennas, a problem that, although seemingly elementary,
is still in search of convincing solutions. Moreover, as mentioned above, this research direction seems to
be in line with the experimental facts.
\par\smallskip


\begin{thebibliography}{99}
\small 

\bibitem{bart} D. F. Bartlett and T. R. Corle, Measuring Maxwell's displacement current inside a capacitor, 
Phys. Rev. Lett., 55-1 (1985), 59--62.
 
\bibitem{born} M. Born and E. Wolf, {\sl Principles of Optics}, Pergamon Press, 1987.

\bibitem{cai} W. Cai, {\sl Computational Methods for Electromagnetic Phenomena: electrostatics in solvation, 
scattering, and electron transport}, Cambridge Univ. Press, 2013.

\bibitem{cheng} D. K. Cheng, {\sl Fundamentals of Engineering Electromagnetics - III Edition},
Addison-Wesley, 1993.

\bibitem{eng2} W. Engelhardt, On the solvability of Maxwell's equations, Ann. Fond. Louis Broglie,
37 (2012), 3--14.

\bibitem{feyn} R. P. Feynman, R. B. Leighton and M. Sands,
{\sl The Feynman Lectures on Physics}, Addison-Wesley, 1963.

\bibitem{fun2} D. Funaro, {\sl  Electromagnetism and the Structure of Matter}, World Scientific, 2008.

\bibitem{fun3} D. Funaro, On the near-field of an antenna and the development of new devices,
arXiv: 1203.1229v1, 2012.

\bibitem{fun4} D. Funaro, From Photons to Atoms - The Electromagnetic Nature of Matter,
arXiv: 1206.3110v1, 2012

\bibitem{debr} D. Funaro, Charging capacitors according to Maxwell's equations: impossible,
Ann. Fond. Louis Broglie, 39 (2014), 75--93.
 
\bibitem{funka} D. Funaro and E. Kashdan, Simulation of electromagnetic scattering with stationary 
or accelerating targets, Int. J. Modern Phys. C, 26-7 (2015), 1--16.

\bibitem{grif} D. J. Griffiths, {\sl Introduction to Electrodynamics - IV Edition},
Pearson, 2013.

\bibitem{hart} T. Hart and P. Birke, {\sl The Poynting Vector Antenna}, BookBaby, 2016.

\bibitem{inan} U. S. Inan and R. A. Marshall, {\sl Numerical Electromagnetics}, Cambridge Univ. Press,
2011.
  
\bibitem{jackson} J. D. Jackson, {\sl Classical Electrodynamics - II Edition},
John Wiley and Sons, 1975.

\bibitem{khol1} A. L. Kholmetskii, O. Missevitch and R. Smirnov-Rueda,
Measurement of propagation velocity of bound electromagnetic fields in near zone,
J. Appl. Phys., 102 (2007), 013529.

\bibitem{khol3} A. L. Kholmetskii, O. Missevitch, R. I. Tsonchev, R. Smirnov-Rueda and A. E. Chubykalo,
Propagation of {EM} field in near and far zones, Progress in Electromagnetics Research M, 22 (2012), 57--62.
 
\bibitem{khol2} O. Missevitch, A. L. Kholmetskii and R. Smirnov-Rueda,
Anomalously small retardation of bound (force) electromagnetic fields in antenna near zone, EPL, 93 (2011),
64004.

\bibitem{monk} P. Monk, {\sl Finite Element Methods for Maxwell's Equations},
Clarendon Press, 2003.

\bibitem{ryl} T. Rylander, A. Bondeson and P. Ingelstrom,
{\sl Computational Electromagnetics - II Edition}, Springer, 2013.
 
\bibitem{cst} 3D Electromagnetic Simulation Software,
CST, https://www.cst.com/

\end{thebibliography}

\end{document}